# Evaluate the Stack Management in Effect Handlers
# using the libseff C Library

ZeHao Yu

School of Informatics, University of Edinburgh

November 27, 2025

# Abstract

Effect handlers are increasingly prominent in modern programming for managing complex computational effects, including concurrency, asynchronous operations, and exception handling, in a modular and flexible manner. Efficient stack management remains a significant challenge for effect handlers due to the dynamic control flow changes they introduce. This paper explores a novel stack management approach using user-level overcommitting within the libseff C library, which leverages virtual memory mechanisms and protection-based lazy allocation combined with signal-driven memory commitment. Our user-level overcommitting implementation dynamically resizes stacks on-demand, improving memory utilization and reducing waste compared to traditional methods.

We rigorously benchmark and evaluate this novel strategy against conventional fixed-size stacks, segmented stacks, and kernel-based overcommitting, using metrics such as context-switch latency, stack expansion efficiency, multi-threaded performance, and robustness under rapid stack growth conditions. Experimental results demonstrate that kernel-based overcommitting achieves an effective balance between performance and flexibility, whereas our user-level implementation, while flexible, incurs additional overheads, highlighting areas for optimization.

This study provides a detailed comparative analysis of various stack management strategies, offering practical recommendations tailored to specific application requirements and operational constraints. Future work will focus on refining user-level overcommitting mechanisms, mitigating non-deterministic behaviors, and expanding benchmark frameworks to include real-world scenarios.

# Table of Contents

# Chapter 1

# Introduction

Effect handlers are an important technology in modern programming that allows programs to manage various computational effects, such as asynchronous operations, concurrency, and exception handling, in a modular way(Ghica et al., 2022; Dolan et al., 2017; Sivaramakrishnan et al., 2021). Effect handlers give developers more fine-grained control over the execution context of their programs, allowing them to customize, handle, and compose different effects. In this regard, libseff is a library focused on effect handlers with a special emphasis on stack management efficiency, optimized for performance critical areas(Alvarez-Picallo et al., 2024; Ghica et al., 2022). However, implementing efficient effect handlers is challenging, especially in stack management, because effect handlers often need to dynamically change control flow, which can quickly exhaust traditional stack resources and affect system performance and stability(Alvarez-Picallo et al., 2024).

Existing stack management methods, such as fixed-size stacks and segmented stacks, have obvious shortcomings(Ma and Zhong, 2023). Fixed-size stacks are simple and fast, but they cannot dynamically adapt to increasing computational demands, and are prone to stack overflow or waste of memory. In contrast, segmented stacks are more flexible in dynamic adjustment, but the high complexity of managing chained segments and the performance penalty caused by frequent adjustment make them less than ideal for performance-critical applications.

This paper investigates an alternative stack management implementation in the libseff C library, exploring the use of virtual memory mechanisms for in-place stack resizing through user-level overcommitting. Leveraging virtual memory and overcommit strategies allows the stack to dynamically and efficiently resize itself only as required, resulting in superior memory utilization, minimized waste, and enhanced flexibility compared to traditional methods. The core objective of our research is to examine whether user-level overcommitting can deliver the necessary balance between memory efficiency, implementation complexity, context-switching performance, and robustness required for efficient effect handler implementations.

We implemented user-level overcommitting in libseff by employing protection-based lazy allocation combined with signal-driven memory commitment techniques. Our approach was rigorously evaluated against multiple existing stack management methods, including fixed-size stacks, segmented stacks, and kernel-based overcommitting strategies. Through comprehensive empirical performance assessments across multiple dimensions such as context-switch latency, stack expansion efficiency, multi-threaded performance, and robustness under rapid stack growth we systematically documented the advantages and limitations of each stack management strategy.

The contributions of this study include:

1. The design and implementation of a novel user-level overcommitting mechanism leveraging virtual memory techniques specifically tailored for libseff.

2. Comprehensive benchmarking and evaluation of multiple stack management strategies across key performance indicators critical to effect handler implementations.

3. A detailed comparative analysis highlighting trade-offs among fixed-size, segmented, kernel-based overcommitting, and user-level overcommitting strategies.

4. Practical recommendations guiding the selection of optimal stack management methods depending on specific application requirements and operational constraints.

The remainder of this paper is structured as follows: Chapter 2 provides background knowledge on effect handlers, continuations, and existing stack management strategies. Chapter 3 details the implementation of various stack management approaches, focusing particularly on our user-level overcommitting mechanism. Chapter 4 presents the evaluation methodology and performance results. Chapter 5 discusses the implications and limitations of our work, and concludes the paper and suggests directions for future research.

# Chapter 2

# Background

## 2.1  Computational Effects and Effect Handlers

Effect handlers are a general programming feature which can be viewed as extremely lightweight composable user-defined operating systems(Chang and Keisler, 1990). They allow the programmer to define, compose, and customise, a swathe of different effectful features within the programming language. For instance, they enable the implementation of concurrency features such as async/await, generators, and lightweight threads(Fisher and Reppy, 2002). Other examples include automatic differentiation, exceptions, probabilistic programming, and reactive programming(Sivaramakrishnan et al., 2021; Ahman et al., 2021).

### 2.1.1  Theoretical Foundations of Computational Effects

Computational effects represent operations that interact with a program's external environment or alter its execution state beyond pure value computation (Bauer and Pretnar, 2015). In the seminal work, Moggi (1991) proposed modeling these diverse effects (such as exceptions, state, nondeterminism, input/output, and continuations) using monads. Under this approach, a computation returning values from a set $A$ is represented as an element of $T(A)$ for a suitable monad $T$.

Building upon this foundation, Plotkin and Power (2003) developed a more concrete characterization through algebraic effects, describing computational effects by: A set of operations representing the sources of effects (effect constructors) An equational theory for these operations that specifies their properties The basic operational intuition is that each computation either returns a value or performs an operation with an outcome that determines a continuation of the computation ([Plotkin and Power, 2001](#)).

For example, using a binary choice operation `choose`, a computation that nondeterministically selects a boolean value could be expressed as:

```
1 choose(return true, return false)
```

Here, the outcome of making the choice is binary: either to continue with the first argument or the second. This algebraic approach provides a more direct way to represent computational effects as specific operations with defined behaviors.

### 2.1.2 From Algebraic Effects to Effect Handlers

While algebraic effects provide a means to describe and create effects, effect handlers extend this framework by providing mechanisms to interpret and manage these effects. As described by [Plotkin and Pretnar (2009)](#), effect handlers serve as deconstructors that capture and process the operations or side effects produced by computations.

The key innovation of effect handlers is their treatment of the remaining computation as a first-class continuation. When an algebraic operation is performed:

1. The computation is suspended

2. Control transfers to the corresponding handler

3. The handler receives both the operation's parameters and the continuation representing the suspended computation

4. The handler can then manipulate this continuation in various ways

Semantically, an effect handler corresponds to a model (which might not be the free model) of the equational theory, and the semantics of computation is given by a unique homomorphism induced by the universal property of the free model ([Plotkin and Pretnar, 2013](#)). This homomorphism translates the original computation into a result where effects have been properly interpreted.

### 2.1.3 Relation to Traditional Exception Handling

Take exceptions as an example. In traditional exception mechanisms, we typically have: An operation to raise an exception (e.g., $\text{raise}_e()$), which serves as an effect constructor, and An operation to handle the exception (e.g., $\text{handle}_e(M, N)$), which acts as the effect handler([Plotkin and Pretnar, 2013](#); [Cabral and Marques, 2007](#)).

The handler not only catches the exception but also determines how to proceed continuing with normal execution if no exception occurs, or switching to an alternative computation when an exception is caught. It is important to note that effect handlers emphasize naturality with respect to evaluation contexts; that is, the operations should behave consistently with the surrounding program context. Additionally, effect handlers aren't just for handling exceptions; they're also useful for handling all kinds of algebraic effects. For example, for effects such as nondeterminism, state management, or input/output, we can arrange the corresponding processor to decide how smoothly to continue execution when these effects occur. Such a unified approach not only makes the theoretical model more straightforward, but also provides greater flexibility for programming language design.

Effect handler is a mechanism to capture and process various operations or side effects while a program is running. It leverages the common features of free models to generate unique mappings, thus providing a unified and flexible semantic framework for all side effects.

### 2.1.4 Applications and Programming Language Support

Effect handlers have become a versatile mechanism for solving a variety of programming problems. They enable the modularization and management of computing effects. Therefore, it has been widely used in research and practical application.

In fact, effect handlers are a simple and effective solution for expressing a variety of programming abstractions. For example, Dolan et al. (2017) shows how effect handlers implement coroutines, async/await patterns, and lightweight concurrency The mechanisms. These examples show that effect handlers can simplify control flow patterns that have traditionally relied on specialized language features or complex library support.

The implementation of effect handler is distributed across ecosystems of different programming languages. For example, in c + + ((Ghica et al., 2022)), OCaml ((Sivaramakrishnan et al., 2021)) and Java ((Brachthäuser et al., 2018)) are the realization of the corresponding, Scala has a similar implementation ((Brachthäuser et al., 2020)). In addition, the research language Eff is designed around algebraic effects and handlers, It serves as an experimental platform to demonstrate the many ways in which computational effects can be defined and combined.

Although effect handlers are very useful, there are still quite a few challenges when it comes to mass adoption in real systems. Ahman et al. (2021) found that large-scale implementations encounter issues of security, modularity, interoperability, readability, and efficiency. Different programming languages implement effect handlers in very different ways, Common methods include including capability-passing style (Brachthäuser et al., 2020), continuation-passing style transformations and abstract machines(Hillerström et al., 2020).

In terms of performance, analysis of Dolan et al. (2017) and Ghica et al. (2022) shows that Implementations of the effect handler can compete in performance with other concurrency mechanisms, although implementation complexity varies.

Now, mainstream languages are increasingly adopting models inspired by effect handlers, Although it is usually not exposed to its full abstraction capabilities. This trend shows that effect handlers are gaining widespread recognition for their usefulness in simplifying the complex control flow of modern programming languages.

Implementation effect handlers, on the other hand, especially in the aspect of management execution context and the stack, facing unique challenges.Since continuations may need to be captured and recovered multiple times, efficient stack management is essential for practical implementation. This is particularly relevant for the libseff C library discussed in this paper, which specifically explores different stack management strategies designed for effect handler implementations.

## 2.2 Continuations and Coroutines

Continuations and coroutines are the central mechanisms to implement effect Handlers, which connect theoretical concepts with practical system designs. These abstractions support the complex control flow patterns required by the effect handler.

### 2.2.1 Continuations

In effect handlers, continuations represent what's left of the program's execution state when the program encounters an effect and temporarily interrupts execution, a continuation is a snapshot of the computation that hasn't been executed yet(Hillerström et al., 2020). Continuations are treated as first-class objects, meaning that they can be passed as arguments to functions, assigned to variables, returned from functions, and manipulated dynamically throughout the program.

When an effect is captured, the handler can resume the paused computation using the resume action. With this mechanism, control returns to the previously captured continuation, and the program can resume execution from where it was interrupted after it has processed the effect. This approach provides advanced functionality, such as sophisticated exception handling, while maintaining logical program flow in effect management.

As a programming concept, a continuation represents the next step of the current computation. The program can capture the current state of execution and resume execution at any time by manipulating these continuations. By "first-class," we mean that continuations can be treated like normal values-passed between functions, stored in data structures, and manipulated by programs.

### 2.2.2   Coroutines

Coroutines are program components that can pause and resume execution, enabling non-blocking concurrent operations(Alvarez-Picallo et al., 2024). Unlike traditional subroutines, coroutines can pause the current state, preserve local variables and execution locations, and then pick up where they left off.They are equivalent to lightweight "threads" that do not rely on the operating system's thread scheduling, but are managed by the programming language or library's internal scheduler.

Coroutines are particularly well suited for handling I/O operations, concurrent tasks, and iterative workflows without blocking other parts of the program. By ceding control at designated locations, they enable cooperative multitasking and avoid the overhead of traditional thread context switches. This property makes them perfect for implementing effect handlers, which need to pause computation during an effect operation and resume execution once processing is complete.

### 2.2.3   Relationship Between Continuations and Coroutines

The relationship between continuations and coroutines forms the basis for effect handlers. continuations capture the state of a program's computations when it encounters Effects, and coroutines provide a mechanism for pausing and resuming these computations in a controlled manner. Together, they provide the infrastructure necessary to implement the flexible control flow patterns that make effect handlers powerful.

When implemented properly, this infrastructure allows programs to perform complex control-flow operations while still maintaining logical coherence while striking a balance between powerful abstractions and practical performance.

## 2.3   Effect Handler C Library: Libseff

As a low-level systems programming language, C generally does not provide high-level control flow features like effect handlers. However, with the rise of effect handlers in programming language theory, several specialized C libraries were developed to fill this gap. Libhandler[1] and libmpeff[2] are two pioneering implementations in the field. However, these libraries mainly serve as runtime support for compiling higher-order languages to C, rather than providing a direct API for C programmers.

Libseff is a library for C programmers that provides the effect handler functionality directly(Alvarez-Picallo et al., 2024). Effect handlers are tools for managing program side effects, such as I/O operations, exception handling, concurrency management, and state modifications, that interact with the external environment or change the program state during computation.

Although libhandler and libmpeff already exist as C effect handlers, Libseff is fundamentally different in its design philosophy. The key difference lies in the target audience: the former is intended primarily for compiler developers as a higher-order language to compile against, whereas Libseff is intended for direct use by C programmers. This means that existing libraries are more suitable for compiler-generated code than for direct use by C developers.

An important feature of Libseff is the use of mutable coroutine objects instead of immutable Continuations. This way, the system doesn't need to create a new continuation every time an effect is executed. By reducing this frequent creation, Libseff greatly reduces the overhead of memory allocation and improves performance, while also reducing memory usage.

### 2.3.1   Runtime Representation

At runtime, each computation with side effects exists as a `seff_coroutine_t` object that maintains the execution state, stack frames, and set of effects that can be processed by the concurrent process. Specifically, each coroutine object contains:

- A status flag (which can be RUNNING, SUSPENDED, or FINISHED) indicating its current activity state

---

[1] Daan Leijen. 2019. *libhandler.* https://github.com/koka-lang/libhandler..

[2] Daan Leijen and KC Sivamarakrishnan. 2023. *libmpeff.* https://github.com/koka-lang/libmprompt..

- A parent concourse pointer that forms a stack-like chain table for quickly locating the appropriate processing context during effect operations

- A 'resume context' that stores execution state information when the coroutine was last resumed or suspended (on x86-64 Linux platforms, this includes instructions, stack data, frame pointer, and all caller-held registers)

- A stack pointer referencing allocated heap memory, which can be either fixed-size or a chained list of 'stacklets'

To optimize effect data delivery, Libseff stores effect data directly on the coroutine's stack and encapsulates it with the effect tag in a `seff_request_t` structure. This structure consists of only two 64-bit fields that can be passed through processor registers, avoiding additional heap memory allocations and data copies. This approach requires programmers to ensure that effect data isn't passed through coprocessor registers during operations and that references to effect data aren't saved beyond the coprocessor's lifecycle or between consecutive calls to `seff_resume`.

### 2.3.2 Primitives

Libseff's low-level implementation includes a set of assembly-written context-switching primitives for coroutine management:

- `seff_resume`: Resumes a coroutine and returns a request object

- `seff_yield`: Suspends the current coroutine and returns control to the context in which it was last resumed (the caller must ensure that the suspended coroutine is either the current coroutine or an ancestor)

- `seff_exit`: Similar to `seff_yield` but marks the coroutine as terminated, eliminating the need to preserve its execution context

Additionally, Libseff provides primitives for finding and selecting handlers:

- `seff_locate_handler`: Traverses the chain of active handlers to find one that can handle the specified effect

- `seff_perform` and `seff_throw`: Similar to `seff_yield` and `seff_exit` respectively, but accept an effect identifier to determine which processor to suspend and call the default processor if no matching processor is found

The high-level PERFORM macro encapsulates `seff_perform`, constructing effect data on the current stack frame and calling the corresponding primitive. Libseff ensures that the coroutine's stack position remains constant, guaranteeing that pointers on the stack remain valid after coroutine suspension.

### 2.3.3 Example for how to use libseff

In order to more intuitively explain the actual use of libseff, here is an example of how to use it in a C program, explaining the functionality of each part of the code and the idea behind libseff's design.

In general, libseff is an effect handler library based on coroutines, which allows users to define, execute, and process custom effects in C language, so as to realize advanced control flow functions such as exceptions, lightweight threads, generators, etc.

The following is a step-by-step explanation of the function of each part of the code and the usage of libseff:

**1. Defining an Effect**

The code begins by using the macro `DEFINE_EFFECT` to define an effect:

```
DEFINE_EFFECT(read_file, 0, char*, { const char* filename; });
```

Purpose: This line defines an effect called `read_file` with a tag (identifier) of 0, a return type of `char*`, and a single parameter `filename` of type `const char*`.

Meaning: In libseff, an effect is like an operation or command that can be "executed" or "called" within a computation. By defining an effect, you prepare for its later use and for writing its handling code.

**2. Using Coroutines (Creating and Resuming Coroutines)**

**a. Creating a Coroutine:**

In the main function, a coroutine is created using `seff_coroutine_new` with the entry function `read_print`:

```
seff_coroutine_t *k = seff_coroutine_new(read_print, NULL);
```

Purpose: `seff_coroutine_new` creates a new coroutine object that encapsulates the function `read_print` and allocates an independent execution stack for it.

Meaning: In libseff, every effectful (effect-utilizing) computation is wrapped into a coroutine. This allows the computation to be suspended and resumed during execution, enabling custom control flow mechanisms.

**b. Executing the Coroutine and Handling Effects:**

The program defines a function `deal_loop` as an event loop for handling effects.

```
1   void deal_loop(seff_coroutine_t *temp) {
2       bool done = false;
3       while (!done){
4           seff_request_t req = seff_resume(temp, NULL, HANDLES(
                read_file));
5           switch (req.effect){
6               CASE_EFFECT(req, read_file, {
7                   char* text = halooooooooooooo ;
8                   req = seff_resume(temp, text, HANDLES(read_file));
9                   break;
10              });
11              CASE_RETURN(req, {
12                  done = true;
13                  break;
14              });
15          }
16      }
17      seff_coroutine_delete(temp);
18  }
```

`seff_resume`: This is the core function in libseff used to resume or start the execution of a coroutine. Its third parameter, `HANDLES(read_file)`, specifies that the current handler is able to handle the `read_file` effect.

`Effect Request Object` (`seff_request_t`): Every time `seff_resume` is called, the coroutine returns an effect request. When the coroutine calls `PERFORM(read_file, "example.txt")` (as will be seen later), the control is temporarily transferred to the handler, and the request object includes the effect identifier.

`Switch-Case Handling`: The `CASE_EFFECT` branch is used to handle an effect request. When it detects that the request is for `read_file`, the branch is taken. In this example, the code simply defines a string (`text = "halooooooooooooo"`) and then calls `seff_resume` again to pass this return value back to the coroutine. The `CASE_RETURN` branch is used for the case when the coroutine returns normally (without performing any effects). In that case, the loop exits.

Meaning: This pattern demonstrates libseff's approach of "decoupling effect interception from handling code." When an effect is performed within the coroutine via `PERFORM`, the coroutine is suspended, and a request is passed to the handler. The handler then examines the effect and provides a result by resuming the coroutine with a corresponding value.

### 3. Performing the Effect

In the `read_print` function, the macro `PERFORM` is used to call an effect:

```
PERFORM(read_file, example.txt )
```

Purpose: Calling this macro constructs an effect request and passes the argument (in this case, the filename `"example.txt"`) to the effect handler. The coroutine is suspended and waits for an external handler to return a result.

Return Value: After the suspended effect is handled and control is resumed, the value is returned (here assigned to `text`), and `read_print` prints the received string.

Meaning: Using `PERFORM` allows the code to express an operation such as "I need to read the content of a file" without directly performing file I/O. The actual implementation of file reading is provided externally by the effect handler, making the code modular and flexible.

**4. Summary**

The overall flow of the program is as follows:

- **Effect Definition:** The effect `read_file` is defined using `DEFINE_EFFECT`, establishing its interface (parameters and return type).

- **Coroutine Creation:** A coroutine is created using `seff_coroutine_new` to wrap the function `read_print` so that it becomes a suspendable computation.

- **Executing the Coroutine and Handling Effects:** The `deal_loop` function starts an event loop that repeatedly calls `seff_resume` to resume the coroutine.

  - After each resume, the returned effect request (generated by a call to `PERFORM`) is examined using a switch statement:

    * If the request is for `read_file`, an appropriate response is provided (here, simply returning a string).

    * If the coroutine returns normally, the loop exits.

- **Resource Cleanup:** After the coroutine finishes execution, the coroutine object is freed using `seff_coroutine_delete`.

This design demonstrates how libseff separates the invocation of an effect from its implementation, allowing the file-reading operation to be redefined (or substituted) easily. Such modularity enables developers to flexibly extend program behavior based on specific requirements.

**Additional Explanation**

Macros and Type Safety: libseff provides several macros (such as `DEFINE_EFFECT`, `PERFORM`, `CASE_EFFECT`, and `CASE_RETURN`) that simplify effect definition and invocation while using compiler checks to ensure the correctness of the types for parameters and return values.

Coroutine Management: By suspending and resuming coroutines with `seff_resume`, libseff enables the implementation of complex control flows (such as asynchronous operations or exception handling) entirely in user space, without relying on operating system thread switching.

## 2.4 Physical Memory-Based Stack Management Strategies

Stack management is one of the core challenges in implementing effect handlers and coroutines. Stack frames provide memory space for function calls while the program is running, and their management is an important part of any coroutine implementation. When designing effect handlers libraries like libseff, the efficiency of stack management directly affects system performance, memory utilization, and reliability. effect handling operations by their nature require frequent saving, restoring, and switching execution contexts. Whenever a coroutine executes an effect operation, the system must save the current execution state and possibly switch to another stack. Therefore, the allocation and management of stack space is particularly important.

In libseff, stack management not only affects memory efficiency, but also directly affects library performance and ease of use.Ideal stack management needs to find a balance in multiple ways: to reduce the memory footprint and runtime stack operation expenses, and ensure that the stack (to prevent overflow) safety, at the same time provide easy to use abstract interface. This section describes the two physics-based stack management methods implemented in libseff: fixed-size stack and segmented stack, analyzes their advantages and disadvantages, and sets the stage for a subsequent discussion of new approaches based on virtual memory.

### 2.4.1 Fixed-Size Stack

The Fixed-Size Stack method is the simplest way to manage the stack because it does not require dynamic resizing of the stack. Simply allocate a fixed-size block of memory when the coroutine is created. This approach is simple to implement and does not introduce additional memory adjustment overhead since the stack size remains constant at runtime.

However, this approach, while simple, has obvious drawbacks. A fixed-size stack may allocate more memory than is needed, resulting in waste. If the allocation is insufficient, it may lead to a stack overflow. In practice, it is very difficult to accurately predict the required memory size. In order to avoid overflow, it is common to conservatively allocate more memory, but this also reduces memory efficiency. Most implementations tend to over-allocate significantly for reliability, but this comes at the expense of memory efficiency.

### 2.4.2   Segmented Stack

Segmented Stack makes first-class continuations easier to work with by using small stack segments that are linked together. When executed, each function first determines whether there is enough space in the current stack segment to fit a new stack frame; If there is not enough space, the "morestack" function is called to allocate a new stack segment. This reduces the risk of stack overflow and dynamically expands the stack according to different memory requirements.

Despite its many advantages, Segmented Stack methods also introduce some previously underappreciated issues, which may lead to memory efficiency and performance degradation. Among them, the "hot-split" problem is a particularly well-studied challenge in Segmented Stack implementation(Ma and Zhong, 2023).

**Hot-Split Problem** The "Hot-Split" problem occurs in tight loops of function calls. In this case, frequent allocation and deallocation of new stack segments significantly degrades performance. Despite the existence of several optimization techniques, this issue remains an ongoing challenge in the implementation of Segmented Stack (Ma and Zhong, 2023).

To mitigate the hot-split problem, libseff employs a doubly-linked list for stack segment management. When a stack segment becomes unnecessary, rather than immediate deallocation, the system retains it in a linked table for potential reuse, thereby avoiding frequent memory allocations. With this optimization, even in worst-case scenarios, function call overhead increases by only a factor of 11 compared to normal function calls. While this might initially appear substantial, the performance impact remains relatively modest in practice because compilers typically inline small functions, eliminating much of the potential overhead(Alvarez-Picallo et al., 2024).

Another challenge with segmented stacks involves their interoperability with standard library functions. Since segmented stacks depend on function-level stack overflow checks, and standard library functions are typically pre-compiled without such checks, unchecked calls may lead to stack overflow or silent memory corruption. While compilers often address this by reserving additional stack space when calling unsupported functions, this approach increases memory consumption(Ma and Zhong, 2023).

Libseff addresses this interoperability issue through the MAKE-SYSCALL-WRAPPER macro, which generates function wrappers that ensure execution switches to the system stack rather than allocating new stack segments when calling such functions. This approach prevents unnecessary memory allocation while maintaining system stability.(Alvarez-Picallo et al., 2024)

### 2.4.3 Summary

In physical memory-based stack management strategies, fixed-size stacks offer simplicity and low runtime overhead but struggle with the fundamental tension between memory waste and overflow risk. Segmented stacks provide flexibility and dynamic growth capabilities but introduce performance overhead, particularly in hot-split scenarios, and require special handling for library interoperability.

These limitations make it necessary to look for other approaches. One idea is to use virtual memory to manage the stack space, dynamically resize the stack when needed, This allows for higher performance and better memory utilization, which will be discussed in more detail later.

## 2.5 Virtual Memory-Based Stack Management and Overcommitting

The previous sections have pointed out that there are limitations to the stack management strategy based on physical memory. Fixed-size stack has simple structure and low runtime overhead, but it is difficult to balance in preventing memory waste and stack overflow. segmented stacks supports dynamic scaling, but this approach incurs an additional performance penalty, especially in "hot-split" scenarios. Because of this, we intend to take an alternative approach: leveraging virtual memory, which enables the stack to be dynamically resized in place.

### 2.5.1   Existing Virtual Memory-Based Stack Implementations

When discussing the stack management strategy based on virtual memory, the imple-
mentation of libmprompt library attracts special attention. Libmprompt implements a
memory management system. In this system, a large virtual memory area is reserved,
and physical memory is dynamically allocated to expand the stack space according to
the needs, so as to support efficient concurrent operations and exception handling.

In this system, each extended stack (gstack) holds execution state, stack frames, and
auxiliary data. It is managed efficiently by caching and global pool (gpool) mechanism.
memory management relies on the operating system's memory mapping interfaces (e.g.,
mmap, mprotect, and madvise). After a large amount of virtual address space is reserved,
pages are submitted dynamically according to runtime requirements to reduce physical
memory allocation overhead. For platforms that do not support overcommitment
or require exponential commitment policies, the system ensures security and high
performance by catching page faults and applying signal handling mechanisms to scale
on demand.

gpool is designed to utilize a large virtual memory region. The area is divided into
several fixed-size blocks, one for each scalable stack slice. blocks are designed with
no-access gaps between them to detect stack overflow. The Gpool metadata is stored
in the first block of the region, which contains an array that tracks the usage of each
stack slice. All stack slices are initially marked as available, and thread-safe indexing
is implemented via spinlocks. At allocation time, stack addresses are calculated from
indices so that the stack slice can be reused after it is deallocated.

This design accounts for stack growth direction (e.g., reversing indices for downward
growth) and leverages the operating system's page protection mechanisms for on-
demand memory commitment. The system efficiently manages thousands of stack
slices, quickly identifying and committing the corresponding pages during page faults,
constituting a high-performance dynamic stack memory management framework.

Although libmprompt is powerful, its implementation process is relatively complex. In
a multithreaded environment, the global stack pool uses spinlocks to protect the free
stack and ensure that data is consistent during concurrent allocation and deallocation
of stack segments. However, in high concurrency situations, such locks can become a
performance bottleneck and even increase the risk of deadlock. Furthermore, each thread
has its own stack cache and deferred deallocation list, which reduces operating system
calls but makes resource management more difficult. Therefore, threads must carefully clean their respective caches when exiting to prevent memory leaks or accidental use of freed resources.

## 2.5.2 Overcommitting as an Ideal Solution

Given these limitations, we tend to develop a more efficient approach to stack management.By using overcommitting, we can get more flexibility in memory allocation, while avoiding the performance loss that segmented stacks can bring and the memory waste caused by fixed-size stacks.

Overcommitting avoids the trade-offs of both previous approaches by leveraging the operating system's virtual memory management capabilities. Rather than allocating a large amount of physical memory upfront or frequently allocating new stack segments, overcommitting reserves a large virtual address space for the stack but only commits physical memory when actually needed. This ensures efficient memory usage without stack overflow risks and avoids the frequent allocation costs inherent in segmented stacks.

By implementing overcommitting in the libseff library, we can automatically and flexibly deal with dynamically changing stack sizes. At the same time, it remains efficient and avoids additional memory allocation and performance penalty due to segmented stacks. Overall, this approach makes optimal use of stack space.

## 2.5.3 Overcommitting Within the Kernel

Virtual memory allocation and lazy allocation are important mechanisms for optimizing resource utilization in modern operating systems. In lazy allocation, memory allocation is deferred until it is actually needed. More specifically, when an application calls the memory allocation function, the kernel does not immediately allocate physical memory, but instead creates an entry in the virtual address space of the process. Tell the application that the memory area is reserved. physical memory allocation occurs only after a page fault is triggered when memory is first accessed. This lazy allocation strategy allows the kernel to provide much more virtual memory than physical memory, because many reserved memory areas may not be actually used.

The kernel employs a commit mechanism to manage memory allocation and usage. The mechanism focuses on two concepts: commit charge (the total amount of memory allocated but not actually used) and commit limit (a limit calculated from RAM, swap space, and kernel configuration parameters such as `vm.overcommit_ratio`).

When a process requests memory, the kernel first updates the commit charge and then checks whether the commit limit has been exceeded. In strict mode (`vm.overcommit_memory=2`), new allocations are allowed only if the commit charge is below the commit limit; In the heuristic mode (`vm.overcommit_memory=0`), the kernel uses historical usage and some judgment algorithm to decide whether to allocate memory; In unconditional mode (`vm.overcommit_memory=1`), however, the kernel usually does not perform a strict check but allows all allocation requests.

The kernel uses the OOM Killer to protect the system if all of the previously delayed allocations are finally used and may run out of RAM and swap space. OOM Killer will according to the process 'memory usage, priority and its impact on the overall system, select a process terminates.

The implementation of such a committing can improve resource utilization and system performance, but in extreme cases it can also unexpectedly terminate critical services. Therefore, it is necessary to take great care in configuring kernel parameters and memory management, and to closely monitor system memory usage to ensure that the system is stable and reliable.

### 2.5.4 The Necessity of User-Level Stack Management

While from the perspective of the operating system, the kernel-level overcommitting really makes a stack management more convenient, but it also has many disadvantages.First, the way the kernel handles this is often opaque and lacks detailed control. When memory does run out, the system may simply call the memory killer to kill the process without handling the error gracefully. Moreover, the strategies for kernel overcommitting vary from one operating system to another, which makes performance and behavior difficult to predict. For example, Linux allows a more generous memory excess commitment by default, while Windows is completely different, This creates inconsistencies and unpredictability in cross-platform applications.

In order to solve these problems, we propose a new user-mode stack management mechanism, which combines the respective advantages of pre-allocation, delayed allocation and overcommitting, aiming to provide a more flexible and efficient stack management strategy. The design and implementation of this new mechanism is described in detail in the next chapter.

## 2.6  Summary

This chapter comprehensively discusses the theoretical basis and practical application of effect handlers and stack management strategies. We started with a brief introduction to the math behind computational effects, a path that started with monadic representation, worked its way through algebraic effects, and eventually to more powerful effect handlers.

Next, we explore the challenges of implementing effect handlers in C. By analyzing the existing libraries and their shortcomings, we introduce libseff, which is designed for C programmers. The library uses a coroutine architecture and has an API designed with ease of use as well as performance in mind.

In addition, the discussion focuses on two traditional stack management approaches: fixed-size stack and segmented stack. Fixed-size stack is simple and efficient, but there is a balance between memory waste and stack overflow risk. The segmented stack can dynamically grow on demand, which is more flexible but can come at a performance penalty in the "hot-split" scenario.

We concluded by introducing virtual memory-based stack management as a promising alternative approach, highlighting how overcommitting strategies might leverage operating system capabilities to achieve both flexibility and performance. The concept of "resizing stacks in place" was presented as a potential solution to the limitations of existing approaches.

Having established this theoretical and technical foundation, the following chapter will present our novel implementation of a virtual memory-based stack management strategy for the libseff library. We will detail the design decisions, implementation challenges, and optimizations employed to create an efficient and flexible stack management mechanism that leverages virtual memory capabilities. This implementation represents the core contribution of our research—a practical solution to the stack management challenges faced in effect handler implementations.

# Chapter 3

# Implementation

In this chapter, we present the implementation details of stack management in the libseff library adopting the user-level overcommit strategy. First, we discuss the concept of user-level overcommit and then explain its implementation process in detail. Subsequently, we discuss the challenges encountered during the implementation and propose corresponding solutions. Finally, we present implementations of other stack management strategies including fixed-size stacks, segmented stacks, and kernel-based overcommitting. These schemes form the basis of our subsequent comparative analysis.

## 3.1   User-level Overcommitting

Overcommitting is a memory allocation policy that allows the total amount of virtual memory committed by the operating system to exceed the actual physical memory and swap space.This approach takes advantage of the on-demand nature of memory allocation, in which physical resources are allocated only when a program actually accesses and uses memory. The basic principle is that most programs do not use all of their allocated memory at the same time. This allows the operating system to allocate virtual memory more freely, improving overall memory utilization.  In the context of libseff, this strategy enables the system to manage a larger number of coroutines, each of which requires independent memory, and is particularly suitable when memory utilization is low.

But overcommitting has its drawbacks.If all allocated Memory is used at the same time, the system may run Out Of physical memory, triggering the out-of-memory (OOM) killer of the kernel to terminate the process, or even cause the system to crash.Therefore, overcommitting must strike a balance between performance optimization and opera-tional risk. Lazy allocation, also known as demand paging, is an important feature of overcommitting. This memory management mechanism dynamically allocates physical memory for virtual memory addresses at runtime, rather than allocating all required physical memory at process startup or thread creation time. Operating systems typically reserve a large chunk of virtual address space for a process or thread as potential memory for stacks or other data structures, but do not immediately allocate physical memory for the entire region. A page fault is triggered when a program accesses an unmapped virtual page for the first time. When the operating system catches a page fault, it allocates the required physical memory to the virtual page, establishes the mapping, and continues program execution.

This strategy greatly improves memory utilization because only the part that is actually accessed is allocated physical memory.Moreover, startup is faster because the system does not have to map the entire address space to physical memory at the beginning, but instead gradually allocates resources as the program runs. However, this mechanism also brings some performance overhead. When the program frequently accesses new memory pages, a large number of page faults will be generated, which will cause system calls and may affect the performance, especially in the scenarios with high real-time requirements.

## 3.2   User-Level Overcommitting Implementation

The implementation of user-level overcommitting mainly relies on three core mechanisms: virtual memory retention, lazy allocation based on protection, and memory commit based on signal. During initialization, the `init_stack_frame` function allocates a large area of virtual memory using `mmap` and sets the appropriate protection flags. This area includes a guard page to detect stack overflow and the main stack area that is initially marked as unreachable (`PROT_NONE`).

To ensure that the signal handler runs properly under stack overflow conditions, a backup signal stack is created via `init_alt_stack`. This dedicated memory provides a safe execution environment for the signal handler, even if there is a problem with the main stack. Next, the system registers a custom `SIGSEGV` signal handler via `sigaction` to intercept memory access violations.

The key step is the `segv_handler` function. When the program tries to access uncommitted stack memory, the processor produces a segmentation fault, which triggers our handler. The handler first checks whether the fault address is within the allocated stack area. If it is determined to be within range, it computes a suitable memory commit bound. Typically, it commits a little more memory than the current fault address to reduce the occurrence of similar faults in the future. After that, the handler calls `mprotect` to change the protection property of the memory from unreachable to read-write (`PROT_READ | PROT_WRITE`), that is, the memory is committed as needed.

This batch submission method greatly reduces the frequency of page faults. Instead of submitting only one page per fault, the system subrenders multiple consecutive pages at a time, thus finding a good balance between efficiency and resource utilization.

During the resource release phase, the `release_stack_frame` function restores the original signal handler configuration and disallows the fallback signal stack. All allocated memory is freed by means of `munmap` when the frame is no longer needed.

See Appendix A for the full implementation code, which shows how virtual memory management, signal handling, and safety-based memory commit nicely mesh together to enable efficient on-demand stack scaling.

## 3.3   Implementation Challenges and Solutions

In implementing our user-level Overcommit approach, we encountered several significant challenges that required careful consideration and proposed innovative solutions.

### 3.3.1   Signal Handler Failure

A major problem is the execution context of the signal handler. When overflow or data corruption occurs in the main stack, the signal handler may not function properly, resulting in undefined behavior or program crashes. To solve this problem, we employ an independent signal stack. By pre-allocating a dedicated memory area, the signal handler can run on it independently of the state of the main stack.

This approach is particularly critical when the main stack is exhausted or compromised. Otherwise, the signal handler will execute on a broken main stack and may not have enough space to complete the necessary processing, causing further problems. By using a separate signal stack, we ensure the security and stability of the signal processing process.

### 3.3.2 High-Frequency Page Fault

Another big challenge stems from the frequent firing of page fault exceptions during rapid memory usage. In theory, each fault should successfully submit the corresponding page through the signal handler, so that the program can continue to run without segmentation fault. However, in high-frequency fault scenarios, frequent context switching and signal processing can sometimes overwhelm the system response, which may cause signal handler reentry issues or stack overflow. Thus triggering segmentation fault or other unstable behavior.

To solve this problem, we implement a batch commit strategy, borrowing ideas from pre-allocation. The traditional pre-allocation method reserves the whole virtual memory area, but does not allocate physical memory immediately, but dynamically commits page when it is actually accessed. Our batch commit policy extends this principle: instead of committing one page per access, we commit multiple consecutive pages at once, depending on the access pattern.

This method greatly reduces the occurrence of fault by predicting future memory accesses. When a page fault occurs, the signal handler not only commits the page of the current fault, but also commits the subsequent pages at the same time and makes them read-write. This reduces the number of possible future page faults in the same region. By reducing the dependence on processor interrupts and context switching, the batch commit strategy improves system performance, especially in scenarios with rapid memory consumption or deep recursion.

## 3.4 Implementation of other Stack Management Strategies

To fully evaluate our user-level overcommit approach, we introduced three different stack management strategies. These implementations are described below to provide a basis for subsequent comparisons.

### 3.4.1 Fixed-Size Stack Implementation

Fixed-size stack comes from libseff and is one of our base baseline. The detailed code can be found in Appendix B. The core function `init_stack_frame` allocates a fixed-size contiguous block of memory via malloc. The function accepts two arguments:

`frame_size`, which specifies the stack size, and a pointer to the stack pointer (`rsp`). If the allocation fails, the program terminates with an error code (-1). Since x86 stacks typically grow downward, this initializes `rsp` to the end of the allocated memory block, that is, `(char *)frame + frame_size`. This ensures that subsequent stack operations expand down from that point. The function returns a pointer to the allocated block as the base address of stack.

### 3.4.2   Segmented Stack Implementation

segmented stack provides a more complex but flexible approach to stack management and is implemented in libseff, which defines two key functions: `init_segment` and `init_stack_frame`, the detailed code can be found in Appendix C. Where `init_segment` allocates and initializes a new stack segment, `init_stack_frame` creates the initial segment, and `init_stack_frame` creates the initial segment. And the stack pointer is configured to point to its end.

The dynamic growth capability of the implementation depends on two key functions: `seff_mem_allocate_frame` and `seff_mem_release_frame`. When the stack runs out of space, `seff_mem_allocate_frame` allocates a new segment, links it to the current segment, and aliges the new stack to the 16-byte boundary while copying the necessary data. Conversely, when the upper-level call is returned, `seff_mem_release_frame` releases the current segment and reverts to the previous segment.

To ensure that threads are initialized correctly, the implementation wraps `pthread_create` to ensure that `seff_mem_thread_init` is invoked when the thread is started to set up the corresponding memory management subsystem.

### 3.4.3   Kernel-Based Overcommitting Implementation

Our kernel-based overcommitting implementation takes advantage of the virtual memory feature built into the operating system. This method relies on the kernel's on-demand allocation mechanism and uses the `MAP_NORESERVE` flag when `mmap` is invoked. In this way, each coroutine reserves a large virtual address space but does not allocate physical memory until it is actually accessed. The detailed code can be found in Appendix D

# Chapter 4

# Performance Evaluation of Stack Management Strategies

In previous chapters, we discussed the theoretical basis of effect handlers in detail and introduced three different stack management schemes in the libseff library.While our analysis has clarified the conceptual differences between fixed-size stacks, segmented stacks, and innovative overcommitting approaches, their performance can only be validated through rigorous real-world testing to inform specific implementation decisions.

This chapter establishes a comprehensive performance evaluation framework for systematically comparing the impact of these stack management policies on key metrics such as efficiency, scalability, and resource utilization in the implementation of effect handler. We will focus on measuring the various metrics that affect libseff performance, and we will test each strategy repeatedly under controlled experimental conditions to show the pros and cons of each strategy.

The benchmark is designed for common use scenarios of effect handler, and the test content covers the efficiency of context switching, the overhead of stack space expansion, the performance of concurrent execution, and the performance under the limit of memory pressure. These data will help us understand the balance between memory utilization, execution speed and operation flexibility, which provides an important reference for strategy selection in different application scenarios.

Through this experimental analysis, we hope to build a set of data-based selection methods, not only to verify our design ideas, but also to deepen the understanding of efficient memory management techniques in the effect handler library.

## 4.1   Experimental Methodology

### 4.1.1   System Configuration

All benchmarks are run on an Intel® Xeon® Gold running Ubuntu 20.04Executed on a 6154 x86-64 processor, the code is compiled using clang 12.0.0 and compiled with standard optimization flags. This hardware and software configuration provides a consistent environment that facilitates the comparison of the relative performance of various stack management strategies.

To ensure consistency, all benchmarks are run under very low system load, and each test is repeated many times, taking the median as the final result to reduce the impact of system noise and fluctuations. At the same time, we monitor the temperature to ensure that performance is not degraded by the temperature control measures.

## 4.2   Context Switching Performance

All the benchmarks in this subsection are single-threaded. We evaluate the efficiency of the system to switch between different execution contexts by measuring the response time of stack switching. This metric is similar to the rate of shift change in a team, where the slower the shift change, the lower the overall efficiency. Efficient context switching is crucial in the implementation of effect Handlers, since the system frequently switches execution contexts while executing and processing effects.

When a task ends or is paused, the system needs to save the current execution state and load the information of the new task, a process called "stack switching". Although context switching involves several steps, the process of saving and restoring stack information directly reflects the efficiency of various stack management strategies. Testing this part of the performance helps us understand:

1. **Overall scheduling efficiency**: If the switching time is too long, it will significantly slow down the response time of the system, especially for applications that require frequent context switching. This is like too much extra overhead during a shift change, affecting overall productivity.

2. **Direct reflection of stack operation efficiency**: During context switching, the system must first save the stack information of the current task before loading the stack of the next task. Different stack implementations -whether fixed-size, segmented, or overcommitting strategies -vary in their efficiency in this regard.

The measured handoff response time directly reflects the overhead introduced by these stack operations.

Our context switching benchmark evaluates both simple and complex scenarios, giving a comprehensive picture of the performance characteristics of each strategy under a variety of operating conditions.

### 4.2.1  Basic Context Switch Latency

To measure the basic performance of different stack management strategies on context switching, we implemented a simple benchmark that isolates the pure context switching overhead. The following code shows an example of our basic context switch implementation using the libseff library.

```
1  void *fn(void *arg) {
2      seff_yield(seff_current_coroutine(), 0, NULL);
3      return NULL;
4  }
5
6  int main(void) {
7      seff_coroutine_t *k = seff_coroutine_new_sized(fn, NULL,
           150*1024);
8      struct timespec start, end;
9      clock_gettime(CLOCK_MONOTONIC, &start);
10     seff_resume_handling_all(k, NULL);
11     seff_resume_handling_all(k, NULL);
12 }
```

This minimal example shows the basic operations involved in context switching. The coroutine function `fn` yields only once and then returns. In the main function, we:

By measuring the time required for these operations, we isolate the basic stack management overhead from context switching without adding additional computational complexity.

Our results are shown in Table 4.1, where there are clear performance differences between different stack management strategies. In these basic context switching tests, both fixed-size stacks and segmented stacks show excellent performance with a response time of approximately 1 μs. However, the response time of kernel-based overcommitting is about 5 μs, which is 5 times slower than the traditional method.

More notably, the response time of user-level overcommitting is 44 μs, which is significantly higher than all other methods. This performance difference indicates an important trade-off: Although Overcommit is more flexible in memory utilization, its implementation has a large impact on response time. The Kernel-level implementation can take advantage of the system's efficient memory mapping mechanism with little additional overhead. However, the user-level implementation has an extra layer of abstraction and dynamic management operations, which leads to a significant increase in latency.

| Stack Strategy | Time (µs) |
|---|---|
| Fixed size stack | 1 |
| Segmented stack | 1 |
| Over-committing(kernel) | 5 |
| Over-committing(user-level) | 44 |

Table 4.1: Average stack switch latency

These results suggest that fixed-size stacks and segmented stacks may be more appropriate for applications with frequent context switching and strict latency requirements, although they have some limitations in terms of memory management flexibility.

More notably, the response time of user-level overcommitting is 44 s, which is significantly higher than all other methods. This performance difference indicates an important trade-off: Although Overcommit is more flexible in memory utilization, its implementation has a large impact on response time. The Kernel-level implementation can take advantage of the system's efficient memory mapping mechanism with little additional overhead. However, the user-level implementation has an extra layer of abstraction and dynamic management operations, which leads to a significant increase in latency. These results suggest that fixed-size stacks and segmented stacks may be more appro- priate for applications with frequent context switching and strict latency requirements, although they have some limitations in terms of memory management flexibility. It should be noted that in this simplest test version, we only focus on the pure stack switching overhead to evaluate the performance of different stack implementations by isolating yield operations in coroutine. However, real-world scenarios are often more complex and involve additional overhead such as resource contention, synchronization requirements, and state management. These complexities often introduce additionalIt should be noted that in this simplest test version, we only focus on the pure stack switching overhead to evaluate the performance of different stack implementations by isolating yield operations in coroutine. However, real-world scenarios are often more complex and involve additional overhead such as resource contention, synchronization requirements, and state management. These complexities often introduce additional latency in context switching, which affects the overall responsiveness of the system. Therefore, we designed a more challenging test variant to simulate these additional loads and to more fully evaluate how different stack implementation strategies perform under realistic operating conditions.

### 4.2.2 Complex Context Switching Test

In this enhanced test, we introduce additional computational effort, synchronization requirements, and state management operations to simulate more complex scenarios. This test allows us to observe the context switching performance of the three stack implementations under different number of iterations.

```
1    DEFINE_EFFECT(complex_yield, 0, void, { });
2    static void* complex_yield_coroutine(void* arg) {
3    int64_t iterations = *(int64_t*)arg;
4    for (int64_t i = 0; i < iterations; i++) {
5
6        volatile int cal = i * 2;
7        PERFORM(complex_yield, 0);
8    }
9    return (void*)(intptr_t)iterations;
10   }
```

This test defines a `complex_yield` effect with no additional data. The coroutine function yields after performing a simple calculation. Each time the effect is processed, the handler performs additional work through a loop. This loop accumulates values, updates a global state variable, and simulates resource races by acquiring and releasing mutex locks. This extra work mimics a more complex workload, where context switch comes with practical computation and synchronization overhead in addition to simple control passing.

The main function initializes a coroutine that executes the complex yield a specified number of times. It uses high-resolution timers to measure the total elapsed time and outputs the result. This design allows us to evaluate how the effect handler, and its associated context switching mechanism, performs under different stack management strategies. The detailed code can be found in Appendix E.

Our results, shown in Table 4.2, show varying performance for different number of iterations. The response time of segmented stacks is significantly faster when the number of iterations is small -only 11 µs for 10 iterations and 50 µs for 100 iterations, compared to 35 µs and 75 µs for fixed-size stacks, respectively. The response time of kernel-based overcommitting is in between, 30 µs and 65 µs, respectively. The results show that segmented stacks may have low per-switch overhead under light load due to its dynamic memory management capabilities.

| Stack Type | 10 iters (µs) | 100 iters (µs) | 10,000 iters (µs) |
|---|---|---|---|
| Fixed Size Stack | 35 | 75 | 3800 |
| Segmented Stack | 11 | 50 | 4000 |
| Over-committing (Kernel) | 30 | 65 | 3700 |
| Over-committing (User) | 110 | 155 | 4000 |

Table 4.2: Complex context switching performance

However, when the iteration counts are increased to 10,000, the performance gap is significantly reduced. Fixed-size stacks take about 3,800 µs, segmented stacks take about 4,000 µs, and kernel-based overcommitting perform slightly better at about 3,700 µs. This convergence phenomenon indicates that the fundamental efficiency of the context switching mechanism is more important than the specific memory management strategy employed under persistent load.

It should be noted that the above data for Overcommitting (30 µs, 65 µs, and 3,700 µs) all refer to kernel-based implementations. In contrast, our user-level overcommitting implementation has higher response times at light load -110 µs at 10 iterations and 155 µs at 100 iterations, while performing similarly at 10,000 iterations (around 4,000 µs). These results demonstrate an important performance feature: both fixed-size stacks and segmented stacks outperform both overcommit implementations in simple context switching tests. But in a more complex scenario with additional workloads, the relative overhead difference decreases when iteration counts are high. kernel-based Overcommit even shows a slight advantage in the persistently high iteration count scenario. This shows that its memory management efficiency can compensate for its higher overhead in each handover under some loads.

## 4.3  Stack Expansion Efficiency

In real applications, the execution state of programs often shows dynamic and unpredictable behavior. Especially in deep recursion or complex concurrency scenarios, the used stack space may exceed the preallocated capacity, triggering dynamic expansion operations. Testing the efficiency of this dynamic scaling helps to understand how fast and robust the system is under memory pressure and to ensure that scaling does not become a bottleneck in overall performance. In other words, the efficiency of the dynamic scaling mechanism directly affects the latency, throughput and resource utilisation of the system, while guaranteeing the normal operation of the program. By measuring the change in response time during scaling, we can evaluate how the entire memory management strategy performs in the face of extreme or unexpected workloads. However, among segmented stacks, fixed size stack, and overcommitting stack strategies, only segmented stacks has a dynamic expansion mechanism. However, as we saw earlier, overcommitting are similar to dynamic scaling by allocating a large amount of virtual memory and mapping the physical memory when it is actually used. If the reserved virtual memory exceeds the total physical memory, then the actual physical stack space will "expand" as the size of the program grows. For this reason, we ignore the fixed size stack in this section, because it does not extend the stack space and will throw an error if you exceed it.

### 4.3.1   Implementation Details

To test the performance of these stack strategies, we set sufficient default sizes for overcommitting and a "fair" default size for segmented stacks in order to observe the expansion cost and compare with overcommitting.The detailed code can be found in Appendix F.

The purpose of this code is to measure the response time when the dynamic expansion operation is triggered after the default stack space is continuously consumed. At the beginning of the program, an effect named `fill_stack` is defined via `DEFINE_EFFECT(fill_stack, 0, void, {})` to signal the end of the recursive call. The function `fill_stack_rec` calls `PERFORM(fill_stack, 0)` when the recursion depth has reached a predetermined maximum, indicating the end of the iteration. The coroutine function `fill_fn` takes care of calling this recursive function, passing in the maximum recursion depth as an argument to simulate a real application where the stack runs out of space due to deep recursion. In the main function, the program first parses the command line arguments to determine the maximum recursion depth (100 by default), then creates an instance of coroutine and passes this argument. Next, call `seff_resume_handling_all(co, NULL)` to start the coroutine, which executes the recursive process until the `fill_stack` effect is triggered.

### 4.3.2   Results and Analysis

The test results are shown in Table 4.3, from which it can be seen that when dynamic expansion is triggered, the average response time of kernel-based overcommitting is about 115 μs, while segmented stacks is about 3700 μs.  This difference is mainly due to different memory management mechanisms.  kernel-based overcommitting allocates a large contiguous virtual memory in advance and maps physical memory only when needed, so that it can be scaled quickly and with low overhead. However, segmented stacks need to allocate and link new stack segments after the current segment is exhausted, which involves additional memory allocation and complex stack pointer management, resulting in a significant rise in latency.  It is worth noting that our user-level Overcommitting implementation has an average response time of about 208 μs.  Although it also employs an on-demand allocation strategy, its performance is worse than that of kernel-based overcommitting, which indicates that the kernel-level implementation can take advantage of system internal optimizations and efficient physical memory mapping, while the user-level approach incurs additional overhead, resulting in slightly higher response time.  Overall, these results reflect the impact of different stack management strategies on response speed in dynamic scaling scenarios.

| Stack Strategy | Time ($\mu$s) |
|---|---|
| Over-committing(kernel) | 115 |
| Over-committing(user-level) | 208 |
| Segmented Stack | 3700 |

Table 4.3: Average time of different stack strategies

## 4.4   Multi-threaded Performance

The design goals of this test are twofold.  First, we verify the correctness of asynchronous operations and nested effect handling in multithreaded environments.  Second, we evaluate the scheduling and effect handling performance of different stack management policies under complex recursive calls and high frequency coroutine handoff.  This benchmark simulates an extreme use case where the coroutine triggers multiple nested asynchronous effects and recursively calls the effect handler function to combine and summarize the results of multiple threads.

This design is able to reveal possible problems during scheduling, state maintenance, and asynchronous execution. It guarantees each stack management strategy to handle the asynchronous operation and effect.By measuring the response time and observing the effect handling behavior under high load, we can compare the operating efficiency of different policies while ensuring that the system is operating correctly.

### 4.4.1   Implementation Details

Our benchmark programs show how a custom effect system can be integrated with POSIX threads to achieve concurrent asynchronous computation, the detailed code can be found in Appendix G. The program defines a `async_op` effect that fires in coroutines. On each iteration of the loop, the program calls `PERFORM(async_op, i)` to pass a number. Effect requests are handled by the recursive function `handleAsyncOpRec`. When the `async_op` effect fires, it calls `concurrentOperation`, combines the current value with the result of the recursive call, and then performs a bitwise XOR on the result. When coroutine returns, the handler gets the return value.

```
1    DEFINE_EFFECT(async_op, 0, void, { int64_t x; });
2    static inline int64_t concurrentOperation(int64_t x, int64_t y)
        {
3        return (x + y) ^ 0xABCDEF;
4    }
5
6
7    static int64_t handleAsyncOpRec(seff_coroutine_t *k) {
8        seff_request_t req = seff_resume(k, NULL, HANDLES(async_op))
            ;
9        switch (req.effect) {
10           CASE_EFFECT(req, async_op, {
11               return concurrentOperation(payload.x,
                    handleAsyncOpRec(k));
12           })
13           CASE_RETURN(req, {
14               return (int64_t) payload.result;
15           })
16       }
17       return -1;
18   }
```

```
19    typedef struct thread_args_t {
20        int64_t iterations;
21        int thread_id;
22    } thread_args_t;
23
24
25    static void* async_loop(void* arg) {
26        thread_args_t* args = (thread_args_t*) arg;
27        for (int64_t i = args->iterations; i > 0; i--) {
28            PERFORM(async_op, i);
29        }
30
31        return (void*)(intptr_t)args->thread_id;
32    }
33    static int64_t run_async(int64_t iterations, int thread_id) {
34        thread_args_t args = {
35            .iterations = iterations,
36            .thread_id = thread_id
37        };
38        seff_coroutine_t *k = seff_coroutine_new(async_loop, &args);
39        int64_t result = handleAsyncOpRec(k);
40        seff_coroutine_delete(k);
41        return result;
42    }
43
44    void* thread_func(void* arg) {
45        thread_args_t* args = (thread_args_t*) arg;
46        int64_t local_result = 0;
47
48        for (int i = 0; i < 1000; i++) {
49            local_result += run_async(args->iterations, args->
                thread_id);
50        }
51        return (void*)(intptr_t)local_result;
52    }
```

The function `run_async` is used to create a coroutine and start the asynchronous process. This function is called repeatedly by each thread to accumulate the result. The main function takes the number of threads and the number of iterations per thread from the command-line argument and then creates the specified number of threads. After all the threads have finished running, the main function joins them and eventually merges and outputs all the results.

## 4.4.2 Results and Analysis

The results are shown in Table 4.4. Our multi-threaded benchmark not only verifies the correctness of asynchronous operations and recursive effect handling in multi-threaded environments, Performance metrics are also reflected. The final merged result embodies the overhead generated by recursive coroutine call, context switching and effect propagation under different stack management strategies. It is important to note that all stack management implementations -fixed-size stacks, segmented stacks, as well as the two overcommitting methods -ended up with consistent results. This result shows that asynchronous operations and effect propagation are handled correctly by each strategy, thus ensuring semantic correctness regardless of the underlying stack implementation.

| Stack Management Strategy | Aggregated Computation Value |
|---|---|
| Segmented Stack | 216540330000 |
| Fixed Size Stack | 216540330000 |
| Over-committing (kernel) | 216540330000 |
| Over-committing (User) | 216540330000 |

Table 4.4: Multi-threaded Asynchronous Operation Results

However, we observed that user-level overcommitting occasionally produces segmentation faults under certain conditions. This reveals a degree of non-determinism in complex scenarios involving high load, multi-threading, and coroutine-based programming. When the on-demand stack mechanism can submit sufficient memory pages in batches promptly to meet coroutine stack extension requirements, the program executes smoothly and produces correct results. However, if stack expansion occurs too rapidly, or if slight differences in thread scheduling arise, or if memory access exceeds the reserved area, the page fault handling mechanism may fail to keep pace. This can trigger multiple faults and eventually lead to segmentation faults. Consequently, the same program may sometimes calculate results successfully and sometimes crash, highlighting the scheduling and memory allocation challenges faced by user-level overcommitting stack management in boundary conditions.

## 4.5 Rapid Stack Growth Evaluation

In order to test[1] the response of each stack management strategy in the case of rapidly running out of stack resources, we designed a computationally intensive algorithm that quickly consumes a large amount of stack memory. Thus, the specific performance of different strategies in memory expansion, switching execution context and maintaining state can be observed.

### 4.5.1 Implementation Details

A complex procedure was used for this evaluation. This program combines forward computation and reverse-mode automatic differentiation to compute an iterative accumulation function and its derivative with respect to the variable *x*. See Appendix H for the implementation code.Two sets of complementary effects are defined in the code: A suite (`e_ap0`, `e_ap1`, `e_ap2`) manages forward computation operations, Such as constant, negation, addition, and multiplication; The other (`r_ap0`, `r_ap1`, `r_ap2`) handles the computation of reverse propagation, based on the `prop_t` structure that stores the value and its derivative pointer.

In the `example` function, the program does the calculation based on a predefined number of iterations. It initializes the global variable *x*to $\{0.5, \ \&dx\}$ (where `dx` starts with 0.0), The accumulated value and previous value are both set to 1.0 (returned by `r_c`). At each iteration, the program executes `r_a(x, r_c(-1.0))`, equivalent to performing an operation on $x + (-1)$, and then applying negation (using `r_n`) to the result,You then update the running total by multiplying `r_m` with the previous value. At the same time, the new value is added to the accumulated result using `r_a`. When all iterations have finished, the final result is stored in the global variable `result`.

The `handle` function recursively handles the reverse propagation effect, computes the derivative based on the captured operation type, and updates the derivative of the associated variable. Meanwhile, the `reverse` function creates a child coroutine that executes the `example` function and then calls `handle` for backpropagation. Finally, the derivative stored in the global variable *x*(accessed via `x.dv`) is displayed. Additionally, the `evaluate` function manages the forward computation effect by sequentially resuming the coroutine to complete the numerical computation.

The implementation clearly shows how complex iterative calculations and reverse automatic differentiation can be combined using the coroutines and effect mechanism to obtain the final sum result and the sensitivity (derivative) with respect to the initial variable *x*.

---

[1]The test script can be found at https://github.com/effect-handlers/libseff/blob/master/tests/ad.c

### 4.5.2 Results and Analysis

The results from the 100-iteration test demonstrate remarkable consistency across all four stack management strategies, as shown in Table 4.5. Each implementation—segmented stacks, fixed-size stacks, kernel-based overcommitting, and user-level overcommitting—calculated results converging to approximately -4.0. The segmented stack implementation exhibited a minor deviation at -3.999958, but this difference is statistically negligible and does not affect the functional correctness of the implementation.

| Stack Strategy | Output (100 iters) |
|---|---|
| Segmented Stack | -3.999958 |
| Fixed Size Stack | -4.000000 |
| Over-committing (kernel) | -4.000000 |
| Over-committing (User) | -4.000000 |

Table 4.5: Outputs of different stack strategies over 100 iterations

The consistent results obtained with these different strategies prove that all implementations correctly handle computations with fast stack consumption while maintaining computational accuracy. Each strategy reliably preserves program state and execution environment, even under complex iterative computations and automatic differentiation operations that cause stack memory to grow rapidly. This consistency not only proves the correctness of the various implementations, but also shows that they are robust to the demands of rapidly expanding stacks.

# Chapter 5

# Conclusion and Discussion

In the paper, we systematically evaluate and compare three different stack management strategies fixed-size stacks, segmented stacks, and Overcommitting and apply them to effect handlers in the libseff C library. We first present the design decisions, show several memory management approaches, and point out the tradeoffs in performance, memory utilization, system stability versus dynamic stack scaling.

We examined several key performance metrics through a series of carefully designed microbenchmark tests. Basic context switching tests showed that in a simple coroutine switching scenario, fixed-size stacks and segmented stacks significantly outperform kernel-based and user-level overcommitting. This is mainly due to the lower stack management overhead of the former two. While kernel-level overcommitting has only moderate performance degradation, our user-level overcommitting implementation introduces higher overhead. This shows that additional layers of abstraction and complex dynamic memory management can significantly affect the system response speed.

When additional computation, synchronization and state saving operations are added, the performance gap between the strategies is significantly reduced. In tests with high number of iterations, say 10,000, kernel-based overcommitting even slightly outperform fixed-size and segmented stacks. This indicates that the flexibility of dynamic memory allocation can partially compensate for the initial overhead in resource-intensive scenarios.

For the test of dynamic stack expansion, we find that segmented stacks has the ability of dynamic expansion, but the delay is high due to frequent allocation and linking of new stack segments. In contrast, kernel-based overcommitting takes advantage of the operating system's efficient memory mapping mechanism to allocate physical memory only when needed, so it is more responsive. Although our user-level overcommitting implementation is less efficient than kernel-level, it still significantly outperforms segmented stacks on dynamic scaling. These results show that the dynamic memory strategy using the efficient page management mechanism of the operating system can provide a more robust and efficient solution for the rapid stack growth.

In addition, we verify the correctness and stability of all policies in a multi-threaded environment for recursive asynchronous effect handling and frequent coroutine operations. Experimental results show that whether fixed-size stacks, segmented stacks, kernel-based overcommitting or user-level overcommitting, Both guarantee the semantic correctness of coroutine execution and effect propagation.

Finally, by combining iterative calculations with a complex test of reverse-mode automatic differentiation, we examine the response of each strategy under dense stack growth conditions. Tests show that segmented stacks and user-level overcommitting are more susceptible to performance impact in this respect, while kernel-based overcommitting can better balance dynamic memory management with low-latency expansion.

In summary, our evaluation shows that no single stack management strategy is suitable for all scenarios. Fixed-size stacks are suitable for use cases with predictable requirements due to their simplicity and low latency. segmented stacks provides dynamic scaling, but with high overhead. However, kernel-based overcommitting can achieve a good balance between memory flexibility and performance in dynamic and heavy load scenarios. Our user-level implementation, while flexible, requires further optimization to match the response speed and efficiency of the kernel-level policy. Future work will be devoted to improving the user-level overcommitting mechanism, exploring hybrid schemes, and extending these results to a wider range of practical application scenarios.

## 5.1 Limitations

### 5.1.1 Commit Granularity and Memory Utilization

In this study, our user-space on-demand paging implementation adopted the batch commit strategy to reduce page fault occurrences. When fault occurs, this changes the protection attribute of multiple pages from `PROT_NONE` to read and write permissions in a single operation. This method effectively reduces the frequency of signal handler calls. However, this approach also has drawbacks: for programs with low actual memory requirements, a large amount of physical memory may be allocated in advance due to access patterns or stack probing mechanism. As a result, the overall memory utilization efficiency is reduced. In other words, the implementation gives up fine-grained control over true on-demand allocation and may waste too much RAM when actual usage is low or partial.

### 5.1.2   Non-deterministic Behavior

In our implementation, we rely on the `SIGSEGV` signal handling and `mprotect` system calls, And that brings a certain amount of unpredictability. The execution of a program is affected by many factors, such as thread scheduling priority, signal transmission time, context switch mode and so on. Especially at times of high concurrency or rapid stack expansion, race conditions and signal handlers being called repeatedly can arise. As a result, sometimes fault is handled correctly, but sometimes segmentation violation is triggered. This instability increases the debugging difficulty, especially in a multi-threaded environment, when multiple execution contexts simultaneously contend for signal handling resources, which may affect the stability of the whole system.

### 5.1.3   Benchmark Limitations

In the performance evaluation, we mainly adopted microbenchmarks to test various parts of on-demand paging mechanism in multi-coroutine and multi-thread scenarios. These tests are effective for measuring individual performance metrics, but comprehensive macrobenchmarks to simulate real workloads and complex production environments are still lacking. Evaluating the overall performance under continuous load, long-term resource usage patterns, and system stability in complex production environments will be an important direction of future work, which will help to more comprehensively verify the feasibility of user-level on-demand paging mechanism in actual large-scale production environments.

## 5.2   Future Work

In order to address the limitations existing in the current implementation, we will focus on the following improvements in future work. These improvements are based on existing designs in memory management and, in particular, on the design pattern of the gpool/gstack scheme in the libmprompt library.

### 5.2.1 Enhancing Commit Granularity and Memory Utilization

The gpool/gstack implementation of libmprompt provides fine management of the virtual memory area by means of smaller stack blocks and dedicated allocation and deallocation mechanisms. One future research direction could be to add an adaptive batch commit strategy to the existing scheme. This strategy can dynamically adjust the number of pages in a commit based on the actual memory usage pattern and access frequency. By using predictive or feed-back driven mechanisms to determine the optimal commit granularity, we can reduce the number of page faults and avoid unnecessary physical allocations, preserving the performance benefits of batch commit. At the same time, the physical memory utilization is greatly improved.

### 5.2.2 Mitigating Non-deterministic Behavior

To address the indeterminacy issues that arise in multithreaded environments, adopting some techniques from the gpool/gstack architecture is a valid future research direction. Specifically, the use of thread-local storage can be strengthened, the implementation of alternate signal stack can be improved, and a more robust synchronization mechanism can be adopted to reduce interference between threads in signal handling. Future releases should focus on improving reentrancy guarantees and thread isolation for signal handlers. In addition, the asynchronous event mechanism provided by modern operating systems can also reduce the dependence on traditional signal handling, thereby reducing the risk of scheduling nondeterministic and race condition.

### 5.2.3 Comprehensive Benchmark Development

A future research could consider extending our performance evaluation framework by adding macro benchmarks that reflect real application scenarios. These assessments will focus on:

1. Measure overall performance, resource utilization, and stability in heavily loaded environments with many concurrent threads and coroutines running simultaneously.

2. Perform prolonged stress tests to evaluate the robustness of the system under conditions of sustained high concurrency.

These extended evaluation works will more comprehensively verify the feasibility of user-level on-demand paging mechanism in the actual large-scale production environment, so as to make up for the shortcomings of the current mainly relying on microbenchmark testing methods.

# Appendix A

# Code for User-Level Overcommitting Implementation

```c
1      #define _GNU_SOURCE
2
3
4  #include <assert.h>
5  #include <stdio.h>
6  #include <stdlib.h>
7  #include <unistd.h>
8  #include <sys/mman.h>
9  #include <stdint.h>
10 #include <signal.h>
11 #include <string.h>
12 #include  seff_mem.h
13
14
15
16 #define DEFAULT_DEFAULT_FRAME_SIZE (150 * 1024)
17
18 #ifndef PAGE_SIZE
19 #define PAGE_SIZE 4096
20 #endif
21
22
23
24
25 #define GUARD_SIZE (PAGE_SIZE)
26
```

```
27
28
29   #include  seff_mem_common.h
30
31
32   static __thread void *g_stack_region = NULL;
33   static __thread size_t g_allowed_size = 0;
34   static __thread size_t g_total_size = 0;
35
36
37   static __thread struct sigaction old_sigsegv_action;
38
39
40   static __thread stack_t g_alt_stack;
41
42
43   static __thread size_t committed_size = 0;
44
45
46   static inline size_t round_up(size_t size) {
47       return (size + PAGE_SIZE - 1) & ~(PAGE_SIZE - 1);
48   }
49
50
51   static void segv_handler(int sig, siginfo_t *si, void *unused) {
52
53       (void)sig; (void)unused;
54       void *addr = si->si_addr;
55       uintptr_t region_start = (uintptr_t)g_stack_region;
56       uintptr_t allowed_start = region_start + GUARD_SIZE;
57       uintptr_t allowed_end = region_start + g_total_size;
58       uintptr_t fault_addr = (uintptr_t)addr;
59
60       if (fault_addr >= allowed_start && fault_addr < allowed_end) {
61
62           uintptr_t current_commit_end = allowed_start +
                   committed_size;
63           if (fault_addr < current_commit_end) {
64
65               fprintf(stderr, Fault_in_already_committed_region\n );
66               exit(1);
67           }
```

```
68
69              size_t commit_size;
70      if (committed_size == 0) {
71
72                  commit_size = PAGE_SIZE;
73      } else {
74
75                  commit_size = committed_size;
76      }
77
78      if (committed_size + commit_size > g_allowed_size) {
79                  commit_size = g_allowed_size - committed_size;
80      }
81
82      uintptr_t commit_start = allowed_start + committed_size;
83
84
85      if (mprotect((void*)commit_start, commit_size, PROT_READ |
            PROT_WRITE) == 0) {
86                  committed_size += commit_size;
87
88                  return;
89      } else {
90                  perror( mprotect_in_segv_handler_failed );
91                  exit(1);
92      }
93
94       }
95
96
97
98
99
100     if (old_sigsegv_action.sa_sigaction) {
101         fprintf(stderr,  Delegating_fault_to_old_handler\n );
102         old_sigsegv_action.sa_sigaction(sig, si, unused);
103     } else {
104         fprintf(stderr,  No_old_handler,_resetting_signal_and_
                raising\n );
105         signal(sig, SIG_DFL);
106         raise(sig);
107     }
```

```
108  }
109
110
111  static void init_alt_stack(void) {
112      size_t alt_stack_size = MINSIGSTKSZ * 2;
113      void *alt_sp = malloc(alt_stack_size);
114      if (!alt_sp) {
115          perror( malloc_for_alt_stack_failed );
116          exit(1);
117      }
118      g_alt_stack.ss_sp = alt_sp;
119      g_alt_stack.ss_size = alt_stack_size;
120      g_alt_stack.ss_flags = 0;
121      if (sigaltstack(&g_alt_stack, NULL) != 0) {
122          perror( sigaltstack_failed );
123          exit(1);
124      }
125  }
126
127
128  static void release_alt_stack(void) {
129      stack_t disable_stack;
130      disable_stack.ss_flags = SS_DISABLE;
131      disable_stack.ss_sp = NULL;
132      disable_stack.ss_size = 0;
133      if (sigaltstack(&disable_stack, NULL) != 0) {
134          perror( disabling_alt_stack_failed );
135      }
136      free(g_alt_stack.ss_sp);
137      g_alt_stack.ss_sp = NULL;
138      g_alt_stack.ss_size = 0;
139  }
140
141
142  void *init_stack_frame(size_t frame_size, char **rsp) {
143
144      g_allowed_size = round_up(frame_size);
145      g_total_size = GUARD_SIZE + g_allowed_size;
146
147      void *region = mmap(NULL, g_total_size,
148                          PROT_NONE,
149                          MAP_PRIVATE | MAP_ANONYMOUS, -1, 0);
```

```
150        if (region == MAP_FAILED) {
151            perror( mmap_failed );
152            exit(1);
153        }
154        g_stack_region = region;
155
156        init_alt_stack();
157
158        struct sigaction sa;
159        memset(&sa, 0, sizeof(sa));
160        sa.sa_flags = SA_SIGINFO | SA_ONSTACK;
161        sa.sa_sigaction = segv_handler;
162        sigemptyset(&sa.sa_mask);
163        if (sigaction(SIGSEGV, &sa, &old_sigsegv_action) != 0) {
164            perror( sigaction_failed );
165            exit(1);
166        }
167
168        committed_size = 0;
169
170        char *initial_sp = (char*)region + g_total_size;
171        initial_sp = (char*)((uintptr_t)initial_sp & ~((uintptr_t)0xF));
172        if (rsp)
173            *rsp = initial_sp;
174        return region;
175    }
176
177
178 void release_stack_frame(void *stack) {
179        (void)stack;
180        sigaction(SIGSEGV, &old_sigsegv_action, NULL);
181        release_alt_stack();
182        if (munmap(g_stack_region, g_total_size) != 0) {
183            perror( munmap_failed );
184        }
185    }
```

# Appendix B

# Code for Fixed-Size Stack Implementation

```c
#include <assert.h>
#include <stdio.h>
#include <stdlib.h>
#include <string.h>
#include <unistd.h>

#include seff_mem.h

// These numbers change between policies
#define DEFAULT_DEFAULT_FRAME_SIZE 150 * 1024
#include seff_mem_common.h

seff_frame_ptr_t init_stack_frame(size_t frame_size, char **rsp) {
    seff_frame_ptr_t frame = malloc(frame_size);

    if (!frame) {
        exit(-1);
    }

#ifndef NDEBUG
    char *ptr = (char *)frame;
    for (int i = 0; i < frame_size; i++) {
        ptr[i] = 0x13;
    }
#endif
```

```
27    *rsp = (char *)frame + frame_size;
28    return frame;
29 }
```

# Appendix C

# Code for Segmented Stack Implementation

```c
#include <assert.h>
#include <stdio.h>
#include <stdlib.h>
#include <string.h>
#include <unistd.h>

#include "seff_mem.h"

// These numbers change between policies
#define DEFAULT_MIN_SEGMENT_SIZE 0
#define DEFAULT_SYSCALL_SEGMENT_SIZE 8 * 1024
#define DEFAULT_DEFAULT_FRAME_SIZE 1024
#include "seff_mem_common.h"

size_t current_stack_top(void);
__asm__( "current_stack_top:
        movq %fs:0x70,%rax;
        ret; " );

seff_stack_segment_t *init_segment(size_t frame_size) {
    size_t overhead = sizeof(seff_stack_segment_t);
    seff_stack_segment_t *segment = malloc(frame_size + overhead);

    if (!segment) {
        exit(-1);
    }
```

```
27
28  #ifndef NDEBUG
29      char *ptr = (char *)segment;
30      for (int i = 0; i < frame_size + overhead; i++) {
31          ptr[i] = 0x13;
32      }
33  #endif
34      segment->prev = NULL;
35      segment->next = NULL;
36      segment->size = frame_size;
37      segment->canary = (void *)0x999999999999;
38
39      return segment;
40  }
41
42  seff_frame_ptr_t init_stack_frame(size_t frame_size, char **rsp) {
43      seff_frame_ptr_t ret = init_segment(frame_size);
44      *rsp = (char *)ret + sizeof(seff_stack_segment_t) + ret->size;
45      return ret;
46  }
47
48  #define ATTRS __attribute__((no_split_stack, visibility( hidden ),
        flatten))
49
50  /* Internal seff_mem functions called by the runtime */
51  ATTRS void *seff_mem_allocate_frame(size_t *frame_size, void *
        old_stack, size_t param_size);
52  ATTRS void *seff_mem_release_frame(void);
53
54  #define LINK(first, second)   \
55      {                          \
56          first->next = second; \
57          second->prev = first; \
58      }
59  void *seff_mem_allocate_frame(size_t *pframe_size, void *old_stack,
        size_t param_size) {
60      size_t frame_size = *pframe_size;
61      seff_stack_segment_t *current = _seff_current_coroutine->
            frame_ptr;
62      assert(current != NULL);
63
64      size_t new_size = frame_size + param_size;
```

```
65        if (new_size < min_segment_size)
66            new_size = min_segment_size;
67
68        seff_stack_segment_t *new_segment;
69        // FIXME: also account for coroutine pause overhead?
70        if (current->next == NULL) {
71            new_segment = init_segment(new_size);
72            LINK(current, new_segment);
73        } else if (current->next->size < frame_size + param_size) {
74            // TODO: Here we could delete the smaller segment
75            new_segment = init_segment(new_size);
76            LINK(new_segment, current->next);
77            LINK(current, new_segment);
78        } else {
79            new_segment = current->next;
80        }
81
82        _seff_current_coroutine->frame_ptr = new_segment;
83        *pframe_size = new_segment->size - param_size;
84
85        /*
86         * Align the returned stack to a 16-byte boundary.
87         * FIXME: in theory this could cause problems because it means
88         * the stack frame will have LESS available space than
                advertised
89         * FIXME: do this properly instead of this garbage
90         */
91        char *new_stack = (char *)(new_segment + 1) + new_segment->size
            - param_size;
92        while (((uintptr_t)new_stack) % 16 != 0) {
93            new_stack -= 1;
94        }
95        memcpy(new_stack, old_stack, param_size);
96        assert(((uintptr_t)new_stack) % 16 == 0);
97
98        return new_stack;
99 }
100
101 void *seff_mem_release_frame(void) {
102     seff_stack_segment_t *current = _seff_current_coroutine->
            frame_ptr;
103     _seff_current_coroutine->frame_ptr = current->prev;
```

```
104      void *stack_top = (char *)current->prev + sizeof(
             seff_stack_segment_t);
105      return stack_top;
106  }
107
108  typedef struct {
109      void *(*start_routine)(void *);
110      void *arg;
111  } seff_mem_start_routine_args;
112
113  extern void seff_mem_thread_init(void);
114  void *seff_mem_start_routine(void *arg) {
115      seff_mem_start_routine_args args = *(seff_mem_start_routine_args
             *)arg;
116      free(arg);
117      seff_mem_thread_init();
118      return args.start_routine(args.arg);
119  }
120
121  extern __attribute__((weak)) int __real_pthread_create(
122      pthread_t *thread, const pthread_attr_t *attr, void *(*
             start_routine)(void *), void *arg);
123  int __wrap_pthread_create(
124      pthread_t *thread, const pthread_attr_t *attr, void *(*
             start_routine)(void *), void *arg) {
125      seff_mem_start_routine_args *args = malloc(sizeof(
             seff_mem_start_routine_args));
126      args->start_routine = start_routine;
127      args->arg = arg;
128      return __real_pthread_create(thread, attr,
             seff_mem_start_routine, args);
129  }
```

# Appendix D

# Code for Kernel-Based Overcommitting Implementation

```c
1   #define _GNU_SOURCE
2
3
4   #include <assert.h>
5   #include <stdio.h>
6   #include <stdlib.h>
7   #include <unistd.h>
8   #include <sys/mman.h>
9   #include <stdint.h>
10  #include <signal.h>
11  #include <string.h>
12  #include seff_mem.h
13
14
15
16  #define DEFAULT_DEFAULT_FRAME_SIZE (150 * 1024)
17
18  #ifndef PAGE_SIZE
19  #define PAGE_SIZE 4096
20  #endif
21
22  #define GUARD_SIZE (PAGE_SIZE)
23
24  #define STACK_EXPANSION_THRESHOLD (64)
25
26  #include seff_mem_common.h
```

```
27
28
29   static inline size_t round_up(size_t size) {
30       return (size + PAGE_SIZE - 1) & ~(PAGE_SIZE - 1);
31   }
32
33
34   static void *g_stack_region = NULL;
35   static size_t g_allowed_size = 0;
36   static size_t g_total_size = 0;
37
38
39   void *init_stack_frame(size_t frame_size, char **rsp) {
40       g_allowed_size = round_up(frame_size);
41       g_total_size = GUARD_SIZE + g_allowed_size;
42
43       void *region = mmap(NULL, g_total_size,
44                           PROT_READ | PROT_WRITE,
45                           MAP_PRIVATE | MAP_ANONYMOUS | MAP_NORESERVE,
                             -1, 0);
46       if (region == MAP_FAILED) {
47           perror( mmap_failed );
48           exit(1);
49       }
50
51       if (mprotect(region, GUARD_SIZE, PROT_NONE) != 0) {
52           perror( mprotect_guard_page_failed );
53           exit(1);
54       }
55       g_stack_region = region;
56       char *initial_sp = (char*)region + g_total_size;
57
58       initial_sp = (char*)((uintptr_t)initial_sp & ~((uintptr_t)0xF));
59       if (rsp)
60           *rsp = initial_sp;
61       return region;
62   }
63
64
65   void release_stack_frame(void *stack) {
66       if (munmap(stack, g_total_size) != 0) {
67           perror( munmap_failed );
```

```
68        }
69    }
```

# Appendix E

# Code for Complex Context Switching Test

```
1
2
3  #include  seff.h
4  #include  mem/seff_mem.h
5  #include <stdio.h>
6  #include <stdlib.h>
7  #include <stdint.h>
8  #include <time.h>
9  #include <inttypes.h>
10 #include <pthread.h>
11
12
13 DEFINE_EFFECT(complex_yield, 0, void, { });
14 static pthread_mutex_t sync_mutex = PTHREAD_MUTEX_INITIALIZER;
15 static volatile int global = 0;
16
17 static inline void extra_work(void) {
18     int sum = 0;
19     for (int i = 0; i < 100; i++) {
20         sum += i;
21     }
22     global += sum;
23 }
24
25 static void* complex_yield_coroutine(void* arg) {
26     int64_t iterations = *(int64_t*)arg;
```

```
27      for (int64_t i = 0; i < iterations; i++) {
28
29          volatile int cal = i * 2;
30          PERFORM(complex_yield, 0);
31      }
32      return (void*)(intptr_t)iterations;
33  }
34
35
36  static int64_t handle_complex_yield_loop(seff_coroutine_t* k) {
37      seff_request_t req = seff_resume(k, NULL, HANDLES(complex_yield)
            );
38      while (req.effect != EFF_ID(return)) {
39          switch (req.effect) {
40              CASE_EFFECT(req, complex_yield, {
41
42                  pthread_mutex_lock(&sync_mutex);
43                  extra_work();
44                  pthread_mutex_unlock(&sync_mutex);
45                  req = seff_resume(k, NULL, HANDLES(complex_yield));
46              })
47          }
48      }
49      return (int64_t)(intptr_t)req.payload;
50  }
51
52  int main(int argc, char** argv) {
53
54      int64_t iterations = (argc < 2) ? 10000 : atoll(argv[1]);
55
56
57      char buffer[8192];
58      setvbuf(stdout, buffer, _IOFBF, sizeof(buffer));
59      seff_coroutine_t* k = seff_coroutine_new(complex_yield_coroutine
            , &iterations);
60
61      int64_t res = handle_complex_yield_loop(k);
62      seff_coroutine_delete(k);
63
64      return 0;
65  }
```

# Appendix F

# Code for Stack Expansion Efficiency

```c
#include  seff.h
#include  mem/seff_mem.h
#include <stdio.h>
#include <stdlib.h>
#include <stdint.h>
#include <time.h>
#include <inttypes.h>
#include <pthread.h>

DEFINE_EFFECT(fill_stack, 0, void, { });

void fill_stack_rec(int depth, int max_depth) {

    char buffer[1024];
    memset(buffer, 0, sizeof(buffer));
    if (depth < max_depth) {
        fill_stack_rec(depth + 1, max_depth);
    } else {

        PERFORM(fill_stack, 0);
    }
}

void *fill_fn(void *arg) {
    int max_depth = *(int*)arg;
    fill_stack_rec(0, max_depth);
```

```
30      return NULL;
31  }
32
33  int main(int argc, char *argv[]) {
34      int max_depth = 100;
35      if (argc > 1) {
36          max_depth = atoi(argv[1]);
37      }
38
39
40      seff_coroutine_t *co = seff_coroutine_new(fill_fn, &max_depth);
41
42      seff_resume_handling_all(co, NULL);
43
44
45
46
47      seff_coroutine_delete(co);
48      return 0;
49  }
```

# Appendix G

# Code for Multi-threaded Performance

```c
#include  seff.h
#include  mem/seff_mem.h
#include  seff_types.h
#include <pthread.h>
#include <stdio.h>
#include <stdlib.h>
#include <stdint.h>

DEFINE_EFFECT(async_op, 0, void, { int64_t x; });

static inline int64_t concurrentOperation(int64_t x, int64_t y) {
    return (x + y) ^ 0xABCDEF;
}

static int64_t handleAsyncOpRec(seff_coroutine_t *k) {
    seff_request_t req = seff_resume(k, NULL, HANDLES(async_op));
    switch (req.effect) {
        CASE_EFFECT(req, async_op, {
            return concurrentOperation(payload.x, handleAsyncOpRec(k
                ));
        })
        CASE_RETURN(req, {
            return (int64_t) payload.result;
        })
    }
    return -1;
```

```
29  }
30
31
32  typedef struct thread_args_t {
33      int64_t iterations;
34      int thread_id;
35  } thread_args_t;
36
37
38  static void* async_loop(void* arg) {
39      thread_args_t* args = (thread_args_t*) arg;
40      for (int64_t i = args->iterations; i > 0; i--) {
41          PERFORM(async_op, i);
42      }
43
44      return (void*)(intptr_t)args->thread_id;
45  }
46
47
48  static int64_t run_async(int64_t iterations, int thread_id) {
49      thread_args_t args = {
50          .iterations = iterations,
51          .thread_id = thread_id
52      };
53      seff_coroutine_t *k = seff_coroutine_new(async_loop, &args);
54      int64_t result = handleAsyncOpRec(k);
55      seff_coroutine_delete(k);
56      return result;
57  }
58
59
60  void* thread_func(void* arg) {
61      thread_args_t* args = (thread_args_t*) arg;
62      int64_t local_result = 0;
63
64      for (int i = 0; i < 1000; i++) {
65          local_result += run_async(args->iterations, args->thread_id)
              ;
66      }
67      return (void*)(intptr_t)local_result;
68  }
69
```

```
70  int main(int argc, char** argv) {
71
72      int num_threads = (argc < 2) ? 4 : atoi(argv[1]);
73      int64_t iterations = (argc < 3) ? 10000 : atoll(argv[2]);
74
75      pthread_t* threads = malloc(num_threads * sizeof(pthread_t));
76      thread_args_t* targs = malloc(num_threads * sizeof(thread_args_t
            ));
77
78
79      for (int i = 0; i < num_threads; i++) {
80          targs[i].iterations = iterations;
81          targs[i].thread_id = i + 1;
82          pthread_create(&threads[i], NULL, thread_func, &targs[i]);
83      }
84
85      int64_t final_result = 0;
86
87      for (int i = 0; i < num_threads; i++) {
88          void* res;
89          pthread_join(threads[i], &res);
90          final_result += (int64_t)(intptr_t)res;
91      }
92
93      free(threads);
94      free(targs);
95
96
97      char buffer[8192];
98      setvbuf(stdout, buffer, _IOFBF, sizeof(buffer));
99      printf( Final_aggregated_result:_%ld\n , final_result);
100     return 0;
101 }
```

# Appendix H

# Code for Rapid Stack Growth Evaluation

```c
#include  seff.h
#include  seff_types.h

#include <stdio.h>

typedef enum { negate_op } op1_t;
typedef enum { add_op, multiply_op } op2_t;

DEFINE_EFFECT(e_ap0, 0, double *, { double value; });
DEFINE_EFFECT(e_ap1, 1, double *, {
    op1_t op;
    double arg1;
});
DEFINE_EFFECT(e_ap2, 2, double *, {
    op2_t op;
    double arg1;
    double arg2;
});

typedef struct {
    double v;
    double *dv;
} prop_t;

effect_set e_smooth = HANDLES(e_ap0) | HANDLES(e_ap1) | HANDLES(
    e_ap2);
```

```
26
27  double e_c(double x) { return *PERFORM(e_ap0, x); }
28  double e_n(double x) { return *PERFORM(e_ap1, negate_op, x); }
29  double e_a(double x, double y) { return *PERFORM(e_ap2, add_op, x, y
        ); }
30  double e_m(double x, double y) { return *PERFORM(e_ap2, multiply_op,
        x, y); }
31
32  DEFINE_EFFECT(r_ap0, 3, prop_t *, { double value; });
33  DEFINE_EFFECT(r_ap1, 4, prop_t *, {
34      op1_t op;
35      prop_t arg1;
36  });
37  DEFINE_EFFECT(r_ap2, 5, prop_t *, {
38      op2_t op;
39      prop_t arg1;
40      prop_t arg2;
41  });
42
43  effect_set r_smooth = HANDLES(r_ap0) | HANDLES(r_ap1) | HANDLES(
        r_ap2);
44
45  prop_t r_c(double x) { return *PERFORM(r_ap0, x); }
46  prop_t r_n(prop_t x) { return *PERFORM(r_ap1, negate_op, x); }
47  prop_t r_a(prop_t x, prop_t y) { return *PERFORM(r_ap2, add_op, x, y
        ); }
48  prop_t r_m(prop_t x, prop_t y) { return *PERFORM(r_ap2, multiply_op,
        x, y); }
49
50  prop_t result;
51  prop_t x;
52
53  void *example(void *args) {
54
55      size_t iters = (size_t)args;
56
57      printf( iters:_%lu\n , iters);
58
59      double dx = 0.0;
60      x = (prop_t){0.5, &dx};
61
62      prop_t acc = r_c(1.0);
```

```
63      prop_t prev = r_c(1.0);
64      for (size_t i = 0; i < iters; i++) {
65          prev = r_m(prev, r_n(r_a(x, r_c(-1.0))));
66          acc = r_a(acc, prev);
67      }
68
69      result = acc;
70
71      return NULL;
72  }
73
74  void handle(seff_coroutine_t *k, prop_t *response) {
75      seff_request_t request = seff_resume(k, response, r_smooth);
76      if (k->state == FINISHED) {
77          *result.dv = 1.0;
78          return;
79      }
80      switch (request.effect) {
81          CASE_EFFECT(request, r_ap0, {
82              double v = e_c(payload.value);
83              double dv = 0.0;
84              prop_t r = ((prop_t){v, &dv});
85              handle(k, &r);
86
87              break;
88          })
89          CASE_EFFECT(request, r_ap1, {
90              double v;
91              switch (payload.op) {
92              case negate_op:
93                  v = e_n(payload.arg1.v);
94                  break;
95              }
96              double dv = 0.0;
97              prop_t r = ((prop_t){v, &dv});
98              handle(k, &r);
99
100             double *dx = payload.arg1.dv;
101             switch (payload.op) {
102             case negate_op:
103                 *dx = e_a(*dx, e_n(dv));
104                 break;
```

```c
105                }
106                break;
107            })
108            CASE_EFFECT(request, r_ap2, {
109                double v;
110                switch (payload.op) {
111                case add_op:
112                    v = e_a(payload.arg1.v, payload.arg2.v);
113                    break;
114                case multiply_op:
115                    v = e_m(payload.arg1.v, payload.arg2.v);
116                    break;
117                }
118                double dv = 0.0;
119                prop_t r = ((prop_t){v, &dv});
120                handle(k, &r);
121
122                double x = payload.arg1.v;
123                double y = payload.arg2.v;
124                double *dx = payload.arg1.dv;
125                double *dy = payload.arg2.dv;
126                switch (payload.op) {
127                case add_op:
128                    *dx = e_a(*dx, dv);
129                    *dy = e_a(*dy, dv);
130                    break;
131                case multiply_op:
132                    *dx = e_a(*dx, e_m(y, dv));
133                    *dy = e_a(*dy, e_m(x, dv));
134                    break;
135                }
136                break;
137            })
138        }
139    }
140
141    void *reverse(void *args) {
142
143        seff_coroutine_t *child = seff_coroutine_new(example, args);
144
145        handle(child, NULL);
146
```

```
147        printf( %lf\n , *x.dv);

148

149        seff_coroutine_delete(child);

150

151        return NULL;

152    }

153

154    void *evaluate(seff_coroutine_t *k, void *args) {

155

156        double value;

157

158        seff_request_t request = seff_resume(k, NULL, e_smooth);

159        while (true) {

160            switch (request.effect) {

161                CASE_EFFECT(request, e_ap0, {

162                    value = payload.value;

163                    request = seff_resume(k, (void *)&value, e_smooth);

164                    break;

165                })

166                CASE_EFFECT(request, e_ap1, {

167                    switch (payload.op) {

168                    case negate_op:

169                        value = -payload.arg1;

170                        break;

171                    }

172                    request = seff_resume(k, (void *)&value, e_smooth);

173                    break;

174                })

175                CASE_EFFECT(request, e_ap2, {

176                    double arg1 = payload.arg1;

177                    double arg2 = payload.arg2;

178                    switch (payload.op) {

179                    case add_op:

180                        value = arg1 + arg2;

181                        break;

182                    case multiply_op:

183                        value = arg1 * arg2;

184                        break;

185                    }

186                    request = seff_resume(k, (void *)&value, e_smooth);

187                    break;

188                })
```

```
189                    CASE_RETURN(request, { return NULL; })
190            }
191        }
192
193        return NULL;
194 }
195
196 int main(int argc, char **argv) {
197        size_t iters = 100;
198        if (argc == 2) {
199            sscanf(argv[1], %lu , &iters);
200        }
201
202        seff_coroutine_t *k = seff_coroutine_new(reverse, (void *)iters)
               ;
203
204        evaluate(k, NULL);
205
206        seff_coroutine_delete(k);
207
208        return 0;
209 }
```